\documentclass[aip,reprint]{revtex4-1}

\usepackage{graphicx}
\usepackage{dcolumn}
\usepackage{bm}
\usepackage{amssymb}
\usepackage{amssymb}
\usepackage[T1]{fontenc}
\usepackage{textcomp}
\usepackage{xcolor}
\usepackage{soul}
\newcommand{\mum}{\mbox{\textmu m}}

\begin{document}

\title{Stimulated-Raman-Adiabatic-Passage mechanism in a magnonic environment }

\author{Q. Wang}
\affiliation{Fachbereich Physik and Landesforschungszentrum OPTIMAS, Technische Universit\"at Kaiserslautern, D-67663 Kaiserslautern, Germany}

\author{T. Br\"acher}
\affiliation{Fachbereich Physik and Landesforschungszentrum OPTIMAS, Technische Universit\"at Kaiserslautern, D-67663 Kaiserslautern, Germany}

\author{M. Fleischhauer}
\affiliation{Fachbereich Physik and Landesforschungszentrum OPTIMAS, Technische Universit\"at Kaiserslautern, D-67663 Kaiserslautern, Germany}

\author{B. Hillebrands}
\affiliation{Fachbereich Physik and Landesforschungszentrum OPTIMAS, Technische Universit\"at Kaiserslautern, D-67663 Kaiserslautern, Germany}

\author{P. Pirro}
\email{Author to whom correspondence should be addressed: ppirro@physik.uni-kl.de}
\affiliation{Fachbereich Physik and Landesforschungszentrum OPTIMAS, Technische Universit\"at Kaiserslautern, D-67663 Kaiserslautern, Germany}

\date{\today}

\begin{abstract}
We discuss the realization of a magnonic version of the STImulated-Raman-Adiabatic-Passage (m-STIRAP) mechanism using micromagnetic simulations. We consider the propagation of magnons in curved magnonic directional couplers. Our results demonstrate that quantum-classical analogy phenomena are accessible in magnonics. Specifically, the inherent advantages of the STIRAP mechanism, associated with dark states, can now be utilized in magnonics. Applications of this effect for future magnonic device functionalities and designs are discussed.
\end{abstract}

\pacs{}
\maketitle

When it comes to finding alternatives to current CMOS based data processing technologies, wave-based technologies have moved into the focus of considerations. They allow for the implementation of a variety of physical principles for the design of devices, such as coherency, superposition and interference, propagation and nonlinearity. In addition, waves have both amplitude and phase, and both can be used for encoding and processing data. 

In the field of photonics wave-based computing has been addressed  since the 1980s [\onlinecite{Feitelson1988}]. However, although much progress has been made, this technology path lacks potential for miniaturization down to CMOS feature sizes due to constrains caused by the used wavelength, which usually is that of infrared light. Furthermore, devices based on nonlinear interactions demand for specific nonlinear media, which are not easily incorporated into high-density, small-size integrated device structures. 

In analogy to photonics, the field of magnonics aims at the use of spin waves and their quanta, the magnons [\onlinecite{Chumak2015, Kruglyk2010, Krawczy2014, Khitun2010}]. Spin waves are the collective excitations of a magnetically ordered solid. Often, we address the field also by the term magnon spintronics, taking into account, that magnons carry spin, and magnonic spin currents characterize a subfield of spintronics. Magnon spintronics encompasses also the conversion between magnonic, i.e. spin, and (spin-polarized) electric currents. The field bears potential for an alternative technology path for wave- and interference-based applications {[\onlinecite{Csaba2017},\onlinecite{Wang2020IDM}]}. It is characterized by excellent scalability [\onlinecite{Liu2018,Che2020,Wintz2016}], easily accessible nonlinearity [\onlinecite{Verba2018,Krivosik2010,Wang2020ring}] and good transport properties [\onlinecite{Kajiwara2010},\onlinecite{Yu2014}]. Recently, several fundamental logical functionalities, such as fundamental sets of logic gates [\onlinecite{Talmelli2020,Fischer2017,Wang2020}], amplifiers [\onlinecite{Verba2018},\onlinecite{Bracher2017},\onlinecite{Gladii2016}], nano-conduits [\onlinecite{Heinz2020},\onlinecite{Wang2019}] and converters [\onlinecite{Rana2019},\onlinecite{Verba2017}] have been demonstrated successfully. Being wave-based, magnon spintronics bears the advantage of the ease of implementation of computation schemes developed in other areas of wave-based phenomena. In doing so, the full scale of useful magnon properties can be made available. Besides excellent wave propagation properties, these encompass specifically the easy controllability by external magnetic, electric and elastic fields [\onlinecite{Chumak2015}]. 

To open up a new direction in the field of coherent wave-based magnonic phenomena, we discuss in this {letter} the implementation of the quantum-classical-analogy magnonic STImulated Raman Adiabatic Passage (m-STIRAP) process as the first example from a wide class of quantum-classical analogues which can, in principle, be realized in any wave-based system. 

The term "quantum-classical analogy"  was coined in the  field of waveguide optics [\onlinecite{Longhi2007,Dreisow2009,Schwartz2007,Liu2017,Block2014}]. There, it was shown that using classical light in optical waveguides allows for the physical realization of analogies to quantum phenomena known from atom and solid state physics. Examples are the waveguide-optical realization of  Bloch oscillations and {Zener tunneling}, Anderson localization, electromagnetically induced transparency [\onlinecite{Liu2017}], and control of quantum mechanical decay {(Zeno dynamics)} [\onlinecite{Biagioni2008}],  to name only a few. 

The process of stimulated Raman adiabatic passage (STIRAP) [\onlinecite{Gaubatz1990,Bergmann1998,Bergmann2019}] is one of the most well-known examples for these analogies. It describes the population transfer between two states via a third, intermediate state which is needed since direct transitions between the two states are forbidden (e.g. because of dipolar selection rules in atomic physics). 

However, this intermediate state can be very lossy [\onlinecite{Vitanov2017,Bergmann2019}] which leads to a strongly damped population transfer if an {\it intuitive} coupling sequence is used which first couples the initial state to the intermediate one and only subsequently couples the then populated intermediate state to the final state. There is, however, another, {\it counter-intuitive} coupling scheme which first couples the empty final and intermediate states and only afterwards introduces a coupling to the initially populated state.  The power of STIRAP relies on the fact that for this counterintuitive way, it is possible to transfer the population via the intermediate state without populating it significantly. This is because the counterintuitive scheme is based on a slow rotation of an adiabatic eigenstate which (ideally) has no overlap with the intermediate state and agrees with the initial and target states at early, respectively, late times. Since the adiabatic eigenstate does not contain the lossy intermediate state it is often termed a dark state. 

Analogous to the concept in photonics [\onlinecite{Longhi2007}], we replace the time coordinate in the original atomic STIRAP process by the spatial coordinate along the propagation direction of a wave, assuming that there are no reflections leading to waves propagating backwards in space.  Thus, in our case, the population transfer happens between three waveguides which take on the role of the three atomic states.
 
\begin{figure}[]
\begin{center}
\scalebox{1}{\includegraphics[width=8.0 cm, clip]{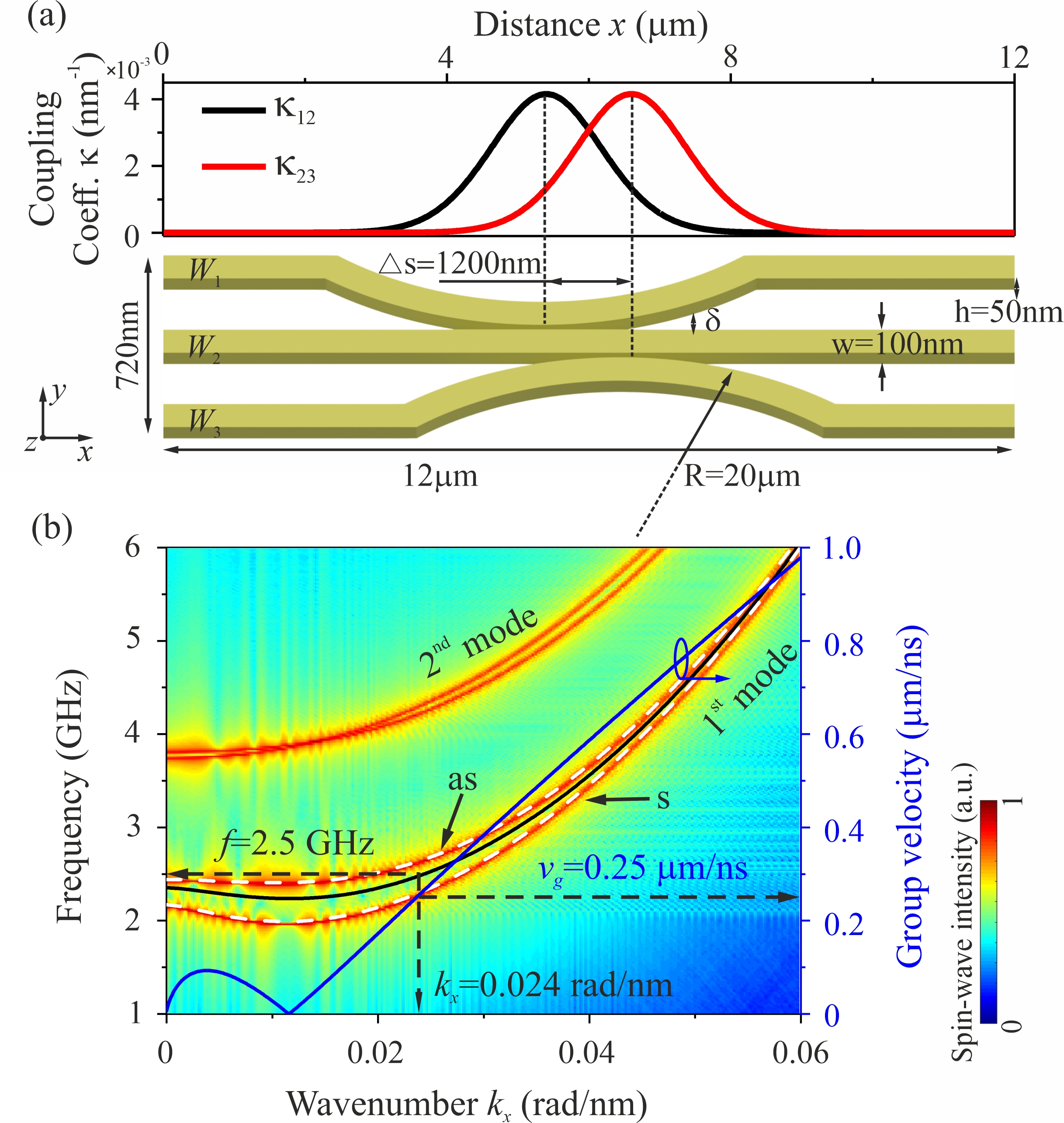}}
\end{center}
\caption{\label{FIG1} {(a) Sketch of m-STIRAP device consisting of three coupled spin-wave nanoscale waveguides. The space dependent coupling coefficients $\kappa_{12}$ and $\kappa_{23}$  are shown in the top panel. (b) The simulated (color map) and calculated (white dashed line) dispersion curves of coupled waveguides with 10 nm gap are shown. The symmetric (s) and antisymmetric (as) modes are marked. The black and blue lines show the calculated dispersion curve and spin-wave group velocity in an isolated waveguide. The assumed parameters of the spin waves used in the simulation and calculation are marked in Fig.\,1(b).}}
\end{figure}

Our demonstrator consists of three, partially curved magnonic waveguides (see Fig.\,1(a)), which locally couple to each other via the dynamic dipolar stray fields of magnons in well defined regions of small separation between two neighboring waveguides. 

To understand the implementation of the m-STIRAP process, we first need to discuss the magnonic directional coupler  [\onlinecite{Wang2018},\onlinecite{Sadvnikov2015}]. The m-STIRAP device consists of two such directional couplers arranged in series and with coherent coupling between them, see Fig.\,1(a). Each coupler  consists of two identical waveguides, which are separated  in transverse direction  by a gap $\delta$, and which is magnetized along the long axis of the waveguides. The dynamic dipolar coupling between the two waveguides leads to the appearance of  two fundamental propagating modes of the coupled system, which, regarding their symmetry, can be categorized as symmetric or antisymmetric with respect to their midplane as shown in Fig.~1(b) [\onlinecite{Wang2018},\onlinecite{Sadvnikov2015}]. 
%the symmetric and the antisymmetric mode [\onlinecite{Wang2018}]. 
Their frequency distance is proportional to the coupling strength. If both modes are excited at the working frequency, the coherent interference of them leads to a periodic beating of the spin-wave energy from one waveguide to the other with a characteristic coupling length $L$. The inverse of this coupling length $L$ is thus a measure of the coupling strength, and we define the coupling coefficient as $\kappa=\frac{\pi}{2L}$. For the parameters relevant here, the dependency of $\kappa$ on the distance $\delta$ between the waveguides is well approximated by an exponential decay [\onlinecite{Wang2018}]  $\kappa \sim e^{-\delta}$. 

Let us turn now back to the actual m-STIRAP device depicted in Fig.\,1(a). The distance $\delta$ between the three waveguides $W_1$, $W_2$, and $W_3$ varies slowly on the scale of the spin-wave wavelength, and so we can assume that the space dependent coupling coefficient $\kappa(x)$ can be locally approximated by the coupling of  two infinite, straight waveguides with the distance $\delta(x)$ between them. Thus, we consider only the coupling between neighboring waveguides. Due to the circular shape of the bending of $W_1$ and $W_3$  and the large bending radius of 20 $\mu$m, the coupling profile has approximately a Gaussian form {(schematically shown in the top panel of Fig.\,1(a))}. From left to right, i.e., in the course of propagation, the point of maximum coupling to the middle waveguide $W_2$ is first reached for $W_1$ and then for $W_3$. Thus, a propagation of spin waves from the left, initially excited in $W_1$ corresponds to the intuitive scheme where spin waves first couple from $W_1$ to $W_2$ and then after the distance $\Delta s$ from $W_2$ to $W_3$.  
%\red{We would like to mention, that, when we remove $W_2$, the coupling between $W_1$ and $W_3$ can be ignored.}

Now we turn to the counter-intuitive scheme. It is realized by initially exciting waves in $W_3$ and considering the output intensity at the end state $W_1$. Here, from left to right, the  coupling between the initial state $W_3$ and the intermediate state $W_2$ reaches its highest value after the maximum coupling between $W_1$ and $W_2$. 
 
We used an analytical theory similar to the matrix Hamiltonian description in [\onlinecite{Longhi2007},\onlinecite{Paspalakis2006}].  We predict the spin-wave amplitudes as a function of distance $x$ from the injection point in a steady state equilibrium using a set of coupled differential equations, neglecting wave reflection:

\begin{eqnarray} 
\label{DGL}
\frac{da_1(x)}{dx}&=&-\beta_1 a_1(x)+i \kappa_{12}(x) a_2(x) \\  \nonumber
\frac{da_2(x)}{dx}&=&-\beta_2 a_2(x)+i (\kappa_{21}(x) a_1(x)+ \kappa_{23}(x) a_3(x)) \\ \nonumber
\frac{da_3(x)}{dx}&=&-\beta_3 a_3(x)+i \kappa_{32}(x) a_2(x)
\end{eqnarray}
 
\noindent where $a_1(x)$, $a_2(x)$ and $a_3(x)$ are the amplitudes of the spin waves in the individual waveguides and $\kappa_{mn} (m,n = 1,2,3)$ is the coupling coefficient between the $m^{\mathrm{th}}$ and $n^{\mathrm{th}}$ waveguide which is related to the coupling length $L_c$ defined in Ref. [\onlinecite{Wang2018},\onlinecite{Sadvnikov2015}] by $\kappa=\frac{\pi}{2L_c}$ for straight waveguides. As mentioned above, the space dependent coupling coefficient $\kappa_{mn}(x)$ can be locally approximated by the coupling of two infinite straight waveguides with the distance $\delta(x)$. Thus it  can be calculated as described in [\onlinecite{Wang2018}]. This implies that both a symmetric and  an antisymmetric mode exist at the chosen working frequency. Their frequency distance is proportional to the coupling strength. Intrinsic losses due to damping are taken into account by the propagation coefficient $\beta_n = 1/L_d (n = 1,2,3)$, where $L_d=v_g \tau$ is the exponential decay length, the product of spin-wave group velocity $v_g$ and lifetime $\tau$. The lifetime itself is inversely proportional to the Gilbert damping constant $\alpha$ of the material.

\begin{figure}[]
\begin{center}
\scalebox{1}{\includegraphics[width=8.0 cm, clip]{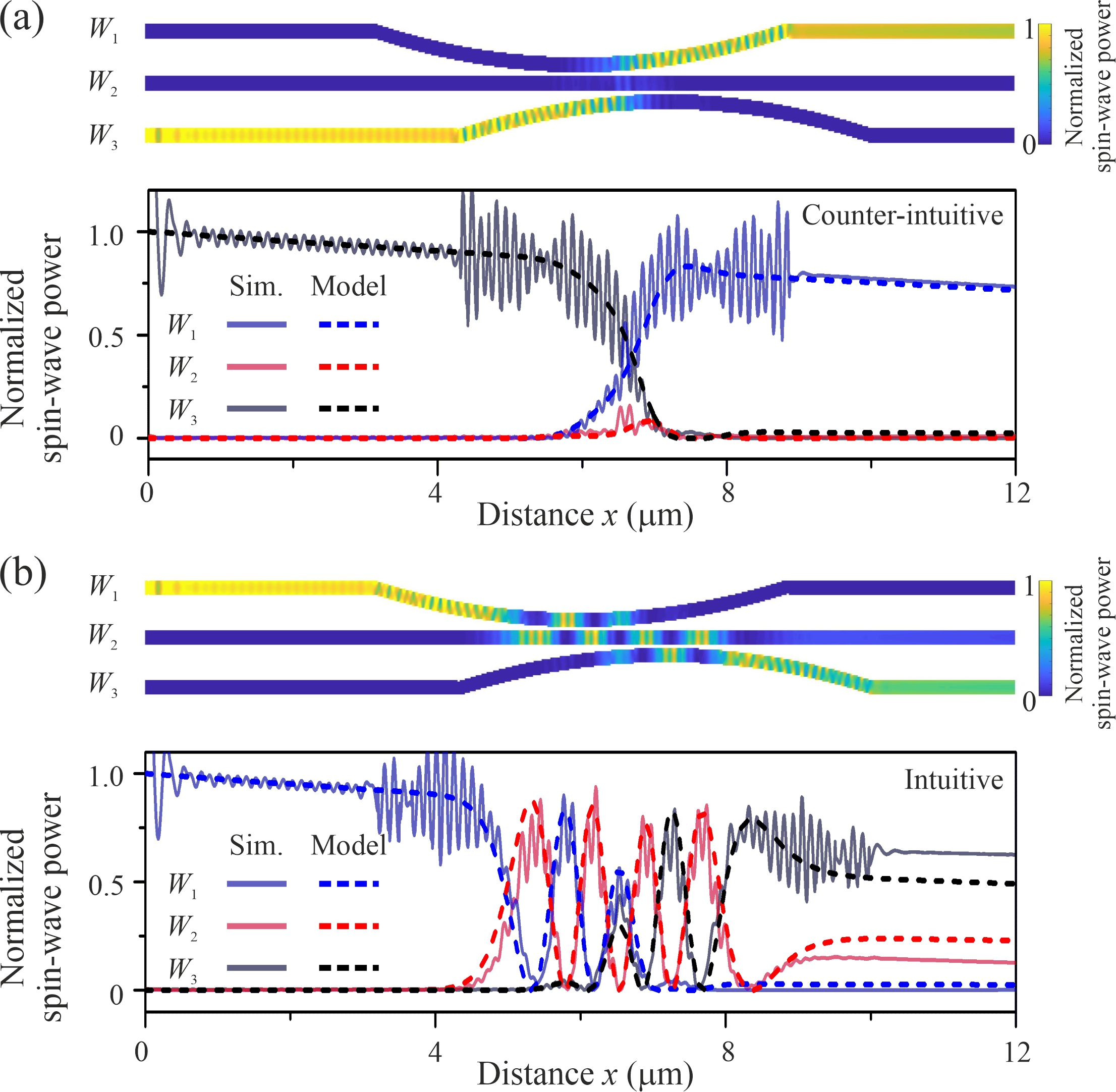}}
\end{center}
\caption{\label{FIG2} Spin-wave propagation for (a) the counter-intuitive scheme, which first couples the two initially non-excited waveguides ($W_1$ and $W_2$) before injecting magnons into the lower one, and (b) the intuitive coupling scheme. 
Shown for each panel is the normalized spin-wave power (proportional to intensity) in a two-dimensional color scheme (upper part) and as a function of position (lower part).
Both numerical simulations (solid lines) and analytical modeling (dashed lines) show that in the counter-intuitive coupling scheme, the excitation of the middle waveguide $W_2$ is very low despite the efficient energy transfer to the output waveguide $W_3$. Thus, this configuration resembles an adiabatic dark state.}
\end{figure}

In the following, if not stated differently, we consider the three waveguides at remanence ($H_\mathrm{ext}=0$) with parallel ground state magnetizations (static $m_x$ components are parallel). As the demonstrator system, we consider magnonic waveguides made of Yttrium Iron Garnet (YIG) with a width of 100 nm and a thickness of 50 nm [\onlinecite{Dubs2020}]. The minimal gap $\delta$ where the waveguides get closest is 10 nm and can be increased by adjusting the frequency of spin waves,  geometry and material of waveguides.  {Figure 1(b) shows the dispersion curves of the first two lowest spin-wave width modes in an isolated single waveguide with the same geometry. The color map was obtained by micromagnetic simulation and the white dashed line was calculated using the analytical model using fully unpinned boundary conditions [\onlinecite{Wang2019}]. A large frequency gap is observed between the first and second width mode. Therefore, in our working frequency range, this waveguide can be treated as a single-mode waveguide. The blue line shows the analytical group velocity of the first width mode. }  Spin waves at a frequency of $f=$2.5 GHz with {a group velocity of 0.25 $\mum$/ns and} a wavelength of around 260 nm are excited {(see Fig.\,1(b))}.

The dashed lines in Fig.\,2(a) show the result of the analytical model for the counter-intuitive coupling case with three identical waveguides where the middle waveguide is almost not populated during the transfer from $W_3$ to $W_1$, thus representing the analogy of adiabatic population transfer through a dark state. The small contribution in $W_2$ is due to non-adiabatic corrections [\onlinecite{Fleischhauer1996}]. We validate our model using a full micromagnetic simulation of the spin wave propagation using Mumax3 [\onlinecite{mumax3}] (solid lines) {with the following parameters: saturation magnetization $\mathrm{M}_s$=1.4$\times10^5$ A/m, exchange constant A=3.5~pJ/m, damping $\alpha_0=2\times10^{-4}$, cell size $10\times10\times50$~$\mathrm{nm}^3$. In order to excite spin waves in the linear regime, a sinusoidal field $b_y=b_0\sin(2 \pi f t)$ was applied to a 20~nm wide area with an oscillation field amplitude $b_0=1$~mT. Outside the bended area, we find a very good agreement of the analytical model and the simulations, whereas inside this area interferences of the initial wave propagating in positive $x$ direction with reflected waves propagation in $-x$ direction are visible in the simulations. Reflected waves occur do to a small impedance mismatch of the spin waves in the straight and in the bended regions due to the slightly changing dispersion relation. Since our analytical model neglects reflections, this standing wave pattern is not reproduced. The reason that there are no standing waves occurring in front of the bended area is that we designed the structure in a way that reflections in this region are suppressed by a coherent destructive superposition of the partial waves reflected from different positions in the bended area similar to the coherent suppression of backscattering in optical microresonators [\onlinecite{Svela2020}].

\begin{figure}[]
\begin{center}
\scalebox{1}{\includegraphics[width=8.0 cm, clip]{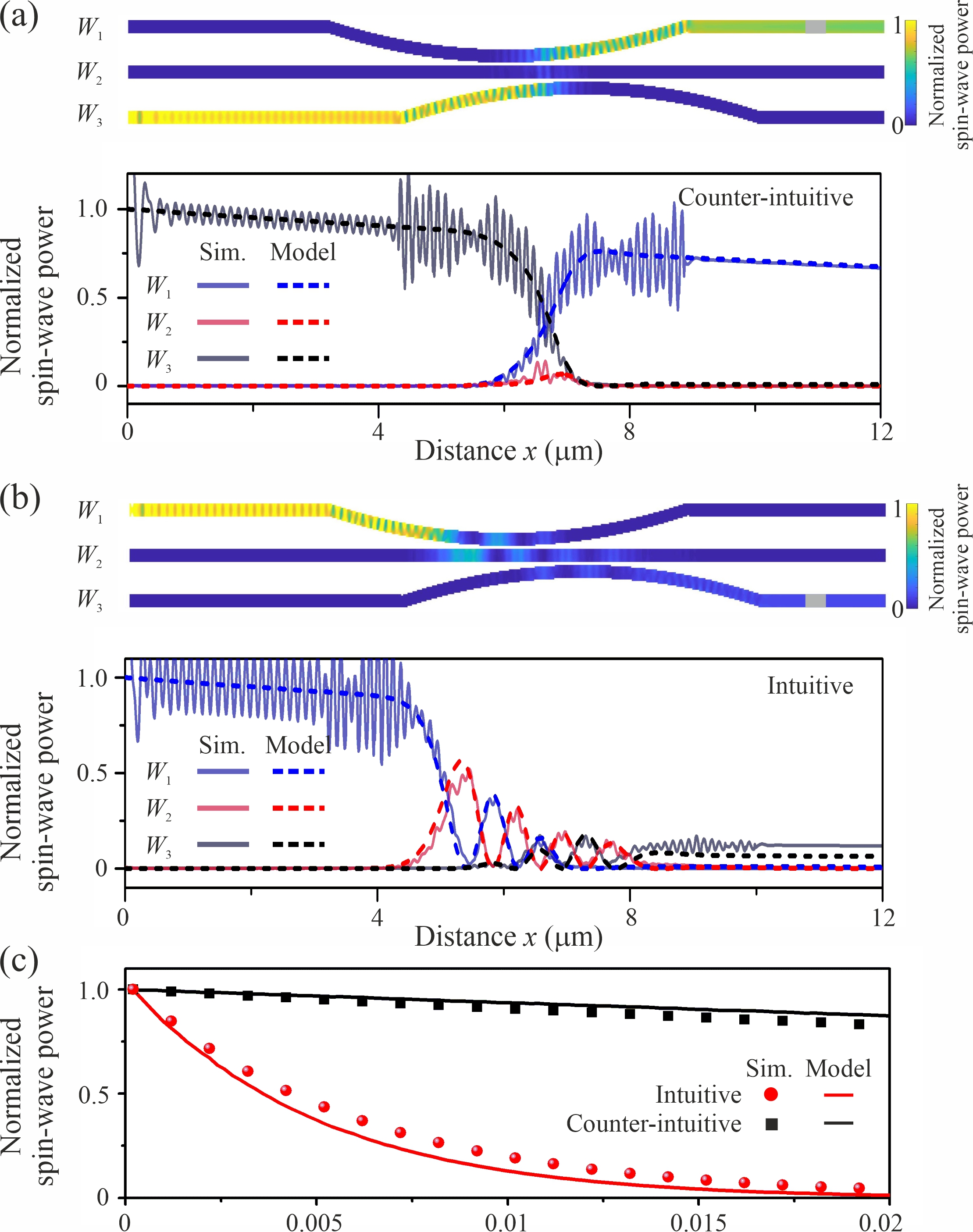}}
\end{center}
\caption{\label{FIG3}(a), (b) Spin-wave propagation similar to Fig.~\ref{FIG2} but with a 50 times increased Gilbert damping parameter in the middle waveguide $W_2$ ($\alpha$($W_2$)=0.01=50 $\times$ $\alpha$($W_{1}, W_{3}$)). A comparison to Fig.~\ref{FIG2}(a)) proves that the adiabatic dark state in the counter-intuitive coupling scheme is not affected by the damping increase whereas in the intuitive coupling scheme, most of the spin-wave energy is dissipated in $W_2$ and only a comparably low part can be transmitted to the output. (c) Output intensity in the counter-intuitive and intuitive coupling scheme as a function of the Gilbert damping parameter $\alpha$ in $W_2$. The read-out of the spin-wave amplitude is performed in the regions marked by the gray rectangles.}
\end{figure}

In contrast to the counter-intuitive case,  $W_2$ is strongly excited for the intuitive coupling case depicted in Fig.\,2(b). This has important implications if the intermediate state has large losses. To demonstrate this, Fig.\,3 shows a scenario similar to Fig.\,2, but with a Gilbert damping constant which is increased by a factor of fifty in $W_2$ only. As it is clearly visible, the counter-intuitive coupling scheme in Fig.\,3(a) is almost not affected since $W_2$ is not significantly excited and, thus, it does not  influence the total dissipation. In contrast, the intuitive coupling scheme shown in Fig.\,3(b) is strongly affected, and only a very low spin-wave amplitude is transferred into the output waveguide $W_3$. Most of the spin-wave energy is dissipated within a short distance in $W_2$. As an additional effect, this leads to a change in the amplitudes of the spin waves reflected from different parts in the bended structure, and, as a result, the complete coherent destructive superposition of the reflected waves is not taking place anymore leading to a strong interference pattern in front of the coupled region. To prove the generality of the observed robustness of the counter-intuitive coupling scheme, the evolution of the intensity in the respective outputs as a function of the Gilbert damping parameter in $W_2$ is shown in Fig 3(c). As expected, it is clearly visible that the damping in $W_2$ in the counter-intuitive coupling scheme is only leading to a very weak drop of the output intensity proportional to $\beta_2 /\Delta s \overline{\kappa}^2$, resulting from non-adiabatic population of waveguide $W_2$ and subsequent losses [\onlinecite{Fleischhauer1996}]. Here $\overline{\kappa}$ is the peak coupling strength inside the coupler. In contrast in the intuitive coupling scheme the intensity decreases exponentially with decreasing life time of the magnons in $W_2$.

A further possibility to check the robustness of the counter-intuitive coupling scheme is the unique possibility of magnonics to reconfigure the coupling strength in between two waveguides by switching from a parallel to an antiparallel alignment of the static magnetizations of the waveguides. As shown in [\onlinecite{Wang2020ring},\onlinecite{Wang2018}], the coupling strength of the antiparallel alignment is enhanced in the presented system by approximately a factor of two compared to the parallel alignment. For the system studied here, this increase of the coupling leads to a splitting of the symmetric and the antisymmetric mode, which is so large that one of them has no solution anymore for the working frequency $f$=2.5 GHz. Thus, the beating between these two modes, which is the basis of the periodic energy exchange between two waveguides, is not possible anymore. Consequently, the requirements of the model described by Eq.\,\ref{DGL} are no longer met. Interestingly, in the counter-intuitive coupling scheme of the adiabatic dark state,  no significant changes are observed even when the analytical model is not valid anymore.  This is evidenced by Fig.\,4(a), where the analytical model and the numerical simulations are shown for this case. 

The surprising robustness of the STIRAP protocol can be understood by the fact that the spin wave is coupled  from $W_3$ via $W_2$ directly into $W_1$. The spin waves in the three waveguides remain in a single adiabatic dark state, which is a coherent superposition of $W_3$ and $W_1$ only, and no periodic energy exchange between the waveguides occurs. One difference to the case of the parallel alignment (Fig.\,2(a)) is that significant reflections leading to a standing wave pattern in front of the bended area occur. The reason is that the changed coupling between the waveguides leads to a change of the impedance mismatch and, consequently, to a change in the relative amplitudes of the waves reflected at different positions in the bended area. Thus, the coherent suppression of backscattering, which has been optimized for the parallel case, is not effective anymore. In contrast, for the intuitive coupling scheme shown in Fig.\,4(b), the increased coupling leads to significant changes compared to Fig.\,2(b) in the spin-wave distribution inside the coupled area as well as in the outputs. The clear disagreement between the analytical model and the micromagnetic simulations shows that the model described by Eq.\,\ref{DGL}, which is predicting a periodic energy exchange between the waveguides, is not longer applicable, as argued above.

\begin{figure}[]
\begin{center}
\scalebox{1}{\includegraphics[width=8.0 cm, clip]{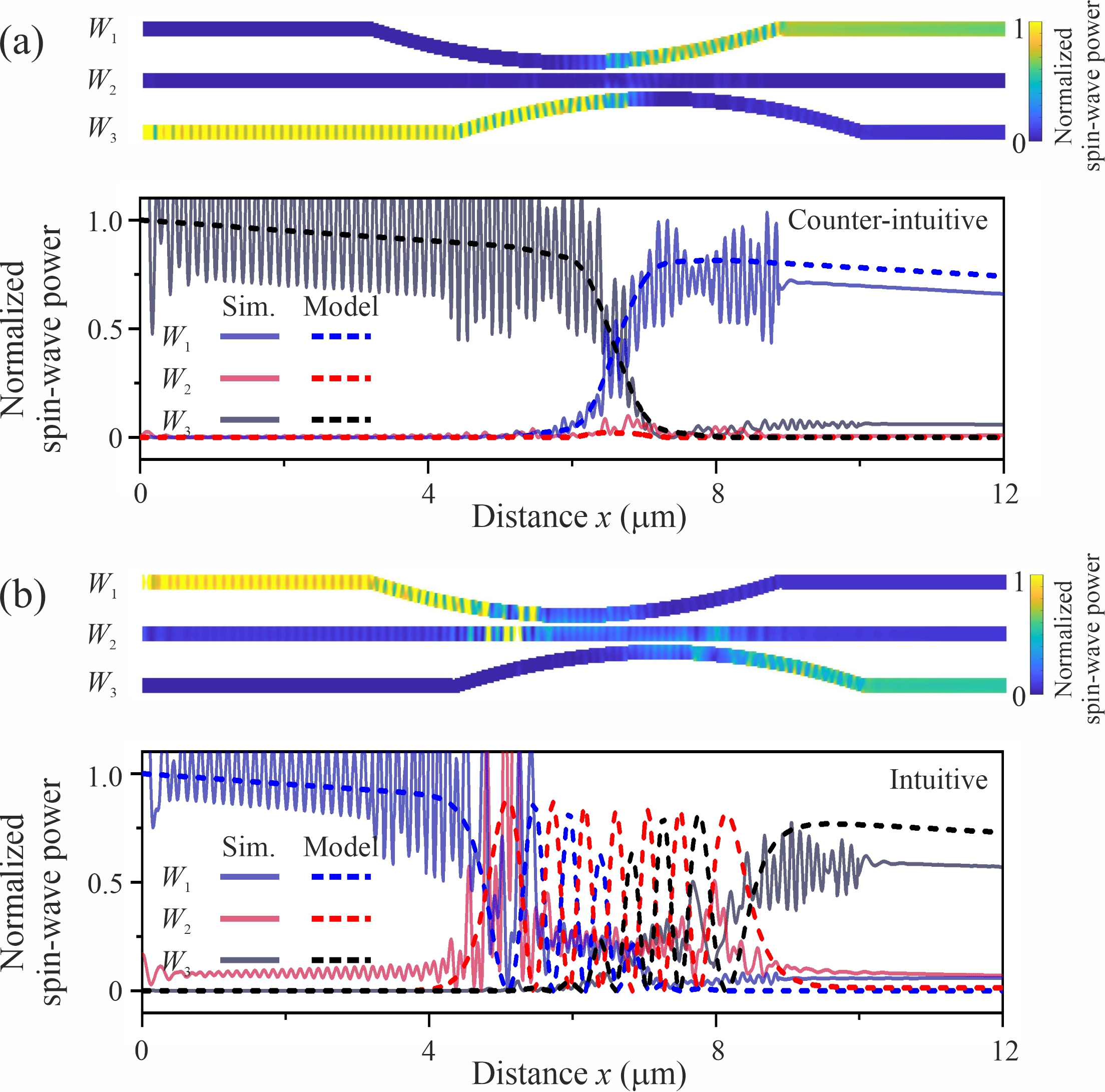}}
\end{center}
\caption{\label{FIG4}Spin-wave propagation similar to Fig.~\ref{FIG2} but with inverted static magnetization in the middle waveguide $W_2$. The increased dipolar coupling does not significantly change the propagation in the counter-intuitive scheme in (a), but the propagation in the intuitive coupling scheme strongly changes  (compare Fig.~\ref{FIG2}b). The disagreement between numerical simulations and the analytical model shows that the requirements of the model used for Eq.\,\ref{DGL} are no longer met.}
\end{figure}

In summary, we have discussed the implementation of the STIRAP process into a magnetic environment using locally coupled, curved magnonic waveguides. We have shown, that the population of magnons can be transferred between two waveguides via an intermediate waveguide, which is not excited, thus their combined wave function resembling the quantum-classical analogy of a dark state. Thus we have transferred a mechanism known from the field of quantum control and coherent control into a magnonic functionality. The proposed scheme to realize magnonic STIRAP can be experimentally realized using state-of-the-art nanopattering techniques developed for nano magnonics [\onlinecite{Heinz2020}]. Challenges for a further development are the implementation of a tunable control of the decay in the inter-mediate state ($W_2$) which can be realized using spintronic effects or parametric spin-wave amplification [\onlinecite{Bracher2017}]. Also, a study of the influence of nonlinear effects is straightforward due to the strong intrinsic nonlinearity of the spin-wave system. In addition, the magnonic system can be time-modulated to enhance the coupling between non-degenerated waveguides, similar to the transition between atomic levels in the original STIRAP concept. We feel that our results  bear high potential for future magnonic device functionalities and designs by bringing together the wealth of quantum-classical analogy phenomena with the wealth of means to control wave propagation in magnonic systems. 

\section*{Data availability}
The data that support the findings of this study are available from the corresponding author upon reasonable request.

\begin{acknowledgments}
The authors thank Andrii Chumak for support and valuable discussions. This research has been supported by the Deutsche Forschungsgemeinschaft (DFG, German Research Foundation) TRR-173 -- 268565370 (Collaborative Research Center SFB/TRR-173 "Spin+X", project B01).
\end{acknowledgments}

\end{document}